\documentclass[preprint,12pt,authoryear]{elsarticle}

\usepackage{amssymb}
\usepackage{amsmath}

\usepackage{hyperref}
\hypersetup{
    colorlinks = true,
    citecolor  = blue,
    linkcolor  = blue,
    urlcolor   = blue,
    breaklinks = true,
    pdfborder={0 0 0},
}
\usepackage{booktabs}
\usepackage{multirow}

\usepackage{enumitem}
\setlist[description]{font=\normalfont\bfseries, leftmargin=0pt, itemsep=0.5ex, labelwidth=0pt}

\usepackage{makecell}
\usepackage{afterpage}

\journal{Nuclear Physics B}

\begin{document}

\begin{frontmatter}



\title{SUMNO: Spatially Unrolled Multigrid Neural Operator for Fluid Flow Surrogates} 


\author[label1,label2]{Yingqi Hu}

\author[label1,label2]{Nan Deng} 

\author[label1,label2]{Changxiao Shao} 

\author[label1,label2]{Xiaozhou He}

\affiliation[label1]{
    organization={College of Artificial Intelligence, Harbin Institute of Technology, Shenzhen},
    postcode={518055},
    country={People's Republic of China}}

\affiliation[label2]{ 
    organization={Institute of Low-Altitude Science and Technology, Harbin Institute of Technology, Shenzhen},
    postcode={518055},
    country={People's Republic of China}}


\begin{abstract}
Data-driven surrogate modeling has emerged as a powerful paradigm for accelerating computational fluid dynamics, and hierarchical U-shaped networks, which resemble the structure of classical Geometric Multigrid (GMG) methods, have become prevalent designs. Modern surrogates built on these designs achieve competitive accuracy at a fraction of the cost of numerical solvers; however, accurate prediction of complex flows remains challenging, and no existing architecture leads across them. In such networks, the multigrid analogy mainly concerns the multi-resolution topology, while their components largely follow common vision practice. To address this, this study introduces \textbf{SUMNO}, a multigrid neural operator for fluid flow fields on regular grids that spatially unrolls a V-cycle into a feed-forward U-shaped network. Parameter-free down- and upsampling stand in for restriction and prolongation, and additive skip connections for coarse-grid correction, with normalization-free transfer and skip pathways. Within this macro-architecture, convolutions are used at the fine and intermediate resolutions and shifted-window attention at the coarsest, whose window partitioning is generalized to handle periodicity independently per axis. Evaluated on unseen test configurations across four unsteady and steady flow benchmarks against five representative baselines, SUMNO is best in 5 of 7 settings and ranks second in the remaining two. It reduces the 10-step rollout MSE by about 30\% relative to the strongest baseline, and outperforms three baselines with over 130 million parameters using 9.4 million parameters.
\end{abstract}


\begin{highlights}
\item Multigrid-aligned macro-architecture for fluid flow surrogates
\item General shifted-window attention operator handling periodicity independently per axis
\item Evaluated on wake, convection, vortex-sheet and cavity flows, unsteady and steady
\item Best on 5 of 7 settings, about 30\% lower rollout error, with 9.4M parameters
\item Parameter-partitioned evaluation protocol and reproducible pipeline
\end{highlights}


\end{frontmatter}

\section{Introduction}

High-fidelity numerical simulations of complex fluid flows, such as direct numerical simulation (DNS) and large eddy simulation (LES), impose prohibitive computational demands due to the broad range of active spatial and temporal scales governed by the Navier--Stokes equations~\citep{Turbulent_Textbook2000}. In response, data-driven surrogate modeling and scientific machine learning (SciML) have emerged as promising alternatives to accelerate computational fluid dynamics (CFD)~\citep{DeepFlow_AIAA2020, MLCFD_NAS2021}. Because discretized Eulerian flow fields on regular grids mirror multi-channel image tensors, adapting architectures from computer vision (CV), from early convolutional networks~\citep{TC_UNet_CMAME2025, CNN4Aero_CM2019} to recent attention-based and hybrid neural operators~\citep{ScOT_NIPS2024, PDETrans_ICML2025, DRIFTNet_ICLR2026, CoSwinNet_Fuel2026}, has become a natural choice. These models achieve competitive accuracy at a fraction of the cost of numerical solvers.

Many leading surrogates adopt a hierarchical U-shaped design, whether as classical convolutional U-Nets~\citep{UNet_MICCAI2015}, attention-based operators such as ScOT~\citep{ScOT_NIPS2024} and PDE Transformer~\citep{PDETrans_ICML2025}, or frequency-coupled models such as DRIFT-Net~\citep{DRIFTNet_ICLR2026}. This layout resembles classical \textit{Geometric Multigrid} (GMG) methods~\citep{MgNet_SCM2019, MGTutorial_Textbook2000, Multigrid_Textbook2000}, in which computation is arranged over a hierarchy of grid levels. The empirical success of these networks underscores the utility of such hierarchical arrangements as an effective structural template for fluid flow modeling. However, in these networks the multigrid analogy mainly concerns the multi-resolution topology, while the components within it, such as resampling, skip fusion and normalization, are largely carried over from vision practice rather than designed around fluid flow fields. Meanwhile, accurate prediction of complex flows, such as cylinder wakes and thermal convection, remains challenging, and the strongest baseline changes from one setting to the next (Table~\ref{tab:main_results}).

Vision practice was developed for natural images, which exhibit substantial redundancy~\citep{ConvNeXt_CVPR2022}. Fluid flow fields are different: they carry spatial differential constraints at every coordinate, where local velocity gradients, strain rates and vorticity, defined by differences between adjacent grid points, dictate the turbulent energy cascade~\citep{Turbulent_Textbook2000}. Fine-scale local continuity therefore has to be preserved at high resolution, whereas patchifying merges neighboring points into one token before any local operation acts on them. Alongside these local constraints, flow fields also carry global couplings, such as the instantaneous pressure coupling between distant vortices in incompressible flow~\citep{FluidDynamics_Textbook2005}. Boundaries differ as well. An image border is an edge, but the boundary conditions of a flow field are set by the geometry of the domain: a periodic axis wraps around, so opposite borders are physical neighbors, and a domain may be periodic along one axis but bounded along the other. Windowed operators designed for images partition the grid with hard borders and do not connect tokens across them~\citep{Swin_ICCV2021}, so such periodic neighbors are separated into unrelated windows, and a single operator cannot serve periodic, partially periodic and bounded domains alike.

Motivated by these insights, this study introduces \textbf{SUMNO} (Spatially Unrolled Multigrid Neural Operator), a multigrid neural operator for fluid flow fields on regular grids. SUMNO spatially unrolls a multigrid V-cycle~\citep{MGTutorial_Textbook2000} into a feed-forward U-shaped network, and its resampling, skip fusion and normalization each follow a multigrid counterpart. Parameter-free downsampling via BlurPool~\citep{BlurPool_ICML2019} and bilinear upsampling stand in for restriction and prolongation; additive skip connections scaled by learnable relaxation factors stand in for coarse-grid correction; normalization is omitted from the transfer and skip pathways; and the network is asymmetric, with a capacity-rich encoder and a slim decoder, reminiscent of the light post-smoothing in multigrid cycles. At the operator level, convolutional blocks at the fine and intermediate resolutions preserve local continuity for gradient modeling~\citep{EarlyConv_NIPS2021, UniFormer_TPAMI2023}, and a shifted-window attention block at the coarsest resolution provides global coupling. The attention operator is generalized to handle periodicity independently per axis (TopoSwin, Section~\ref{subsec:boundary_mechanisms}), so that tokens on opposite sides of a periodic border fall into the same window. Together with circular or reflection padding in convolutions, this allows a single network to cover doubly periodic, cylindrical and non-periodic domains.

SUMNO is evaluated on four benchmark flows: three unsteady dynamical systems, namely the doubly periodic Navier--Stokes sinusoidal vortex sheet (NS-SVS)~\citep{ScOT_NIPS2024}, the open-boundary wake of actuated fluidic pinballs (Pinball)~\citep{Pinball_CSF2025}, and horizontally periodic Rayleigh--B\'{e}nard convection (RBC)~\citep{WellRBC_NIPS2024}, and one steady boundary-value problem, the lid-driven cavity flow with internal obstacles (LDC)~\citep{LDC_CE2025}. Models are trained on single-step transitions and evaluated by autoregressive rollout at horizons of 1 and 14 steps on NS-SVS and 1 and 10 steps on Pinball and RBC, with LDC at steady state, which gives seven settings in total. Under a parameter-partitioned protocol (Table~\ref{tab:dataset_summary}), the training, validation and test sets are disjoint at the trajectory level (the geometry level for LDC), with held-out initial vorticity fields, actuation controls, initial buoyancy fields or obstacle geometries.

On the two complex flows, Pinball and RBC, SUMNO is best in all four settings under both metrics: it roughly halves the single-step MSE of the strongest baseline and reduces the 10-step rollout MSE by about 30\% on both (Table~\ref{tab:main_results}). With 9.4M parameters, SUMNO outperforms ScOT-B~\citep{ScOT_NIPS2024}, PDET-B~\citep{PDETrans_ICML2025} and DRIFT-Net~\citep{DRIFTNet_ICLR2026}, each with over 130M parameters, in all seven settings (Table~\ref{tab:macs_params}). SUMNO is also best on NS-SVS at one step, and ranks second on the remaining two settings, NS-SVS at 14 steps and LDC. Qualitatively, the baselines' errors appear as grid-like textures, fine-scale ripples or smoothed fine structures, whereas SUMNO produces milder errors and retains more fine-scale structure (Fig.~\ref{fig:qualitative_comparison}).

The remainder of this paper is organized as follows: Section~\ref{sec:method} formalizes the multigrid-inspired formulation and details the architectural components; Section~\ref{sec:experiments} outlines the benchmark configurations, baselines, model complexity and training specifications; Section~\ref{sec:results} presents quantitative accuracy evaluations, a step-wise analysis of rollout errors, and qualitative comparisons of the predicted flow fields and their error distributions; and Section~\ref{sec:conclusion} concludes the paper.

\section{Related Work}

\subsection{Neural Surrogates for Fluid Dynamics}
Data-driven surrogates learn the mapping from the current flow state to the next state or to a steady solution, a formulation known as operator learning~\citep{UniversalApprox_IEEE1995, DeepONet_NMI2021, NOP_JMLR2023}. Existing models differ in their overall layout and in the local operator used to mix spatial information.

Early surrogates applied convolutional networks to regular-grid fields, mostly within U-shaped encoder--decoder hierarchies~\citep{UNet_MICCAI2015, BayesConv_JCP2018}, for steady flow prediction, aerodynamics, and time-dependent simulation~\citep{CNN4Steady_KDD2016, DeepFlow_AIAA2020, CNN4Aero_CM2019, LDC_CE2025, PDEAR_JCP2020, MLCFD_NAS2021, TurbSimu_ICLR2022}. Diffusion-based refinement has been used to improve long rollouts of such backbones~\citep{PDERefiner_NeurIPS2023}. Fourier Neural Operators (FNO)~\citep{FNO_ICLR2021} instead stack spectral convolutions without a multi-resolution hierarchy, parameterizing integral kernels in the Fourier domain; they are resolution-independent, and extensions address irregular geometries and non-periodic domains~\citep{GeoFNO_JMLR2023, WNO_CMAME2023}. In these models, spatial information is mixed by convolutional or spectral operators.

A second line adopts attention as the local operator. Galerkin Transformer~\citep{Galerkin_Transformer_NIPS2021} applied self-attention to operator learning, interpreting a linearized variant as a Petrov--Galerkin projection. Subsequent designs use attention to handle irregular meshes, arbitrary query locations, or multiple input functions, making few assumptions on the sampling grid~\citep{LOCA_JMLR2022, OFormer_TMLR2023, GNOT_ICML2023}. For regular grids, ScOT~\citep{ScOT_NIPS2024} and PDE Transformer~\citep{PDETrans_ICML2025} follow vision transformers~\citep{ViT_ICLR2021}, returning to a U-shaped hierarchy and using shifted-window attention (Swin)~\citep{Swin_ICCV2021} as the operator at all resolutions, and are trained at scale on diverse datasets. DRIFT-Net~\citep{DRIFTNet_ICLR2026} retains the ScOT backbone but replaces window attention with parallel convolutional, MLP, and Fourier-domain branches. Thus, recent models share a U-shaped hierarchy and differ in the local operator and training scale. SUMNO adopts the same U-shaped template, but its inter-grid transfer, skip fusion and normalization follow multigrid counterparts (Section~\ref{sec:method}), with convolutions at fine and intermediate resolutions and shifted-window attention only at the coarsest scale. Unlike resolution-independent or mesh-flexible operators, SUMNO targets fixed regular grids.

\subsection{Multiscale Hierarchies and Multigrid}
The U-shaped hierarchy resembles the structure of classical Geometric Multigrid (GMG) methods~\citep{MGTutorial_Textbook2000, MGApplication_Textbook2013, Multigrid_Textbook2000}, in which smoothing on fine grids is combined with restriction, coarse-grid correction, and prolongation. This correspondence has been made explicit in deep learning: MgNet~\citep{MgNet_SCM2019} maps residual convolutional networks to multigrid cycles for elliptic PDEs, and subsequent studies learn relaxation, restriction and prolongation operators or neural preconditioners to accelerate linear solvers~\citep{MgPDE_ICML2019, DNNMG_JCP2022, MetaMgNet_JCP2022}. These works target solvers, whereas SUMNO uses the multigrid cycle as the macro-architecture of a surrogate for flow fields, with parameter-free inter-grid transfer in place of learned transfer operators.

\subsection{Boundary Handling and Topology}
Grid-based networks handle boundaries in two places: convolutions and windowed attention. In convolutions, padding determines how fields are extended beyond the domain, with zero, reflection, and circular padding being common choices; circular padding corresponds to periodic domains and underlies the equivalence between convolution and Fourier-domain filtering~\citep{PDEArena_TMLR2023}, and is employed for periodic flows~\citep{APEBench_NIPS2024}.

In window-based Transformers, Swin~\citep{Swin_ICCV2021} shifts windows by cyclically rolling the feature map and masks attention across the rolled seam. PDE Transformer~\citep{PDETrans_ICML2025} supports fully periodic and fully non-periodic domains. In this work, the mask is derived from the periodicity of each axis rather than selected from these two cases, which recovers both and further covers mixed configurations such as cylinders.

\subsection{Benchmarks and Evaluation Protocols}
Several benchmark collections have been proposed for learning PDE solutions~\citep{PDEBench_NIPS2022, WellRBC_NIPS2024, APEBench_NIPS2024}, offering a wide range of metrics, from RMSE and its normalized or variance-scaled variants to maximum error, conservation, Fourier-band, $H^1$ and correlation-based measures.

Recent models are likewise evaluated with different metrics. ScOT~\citep{ScOT_NIPS2024} and DRIFT-Net~\citep{DRIFTNet_ICLR2026} report the relative $L_1$ error at the final time, whereas PDE Transformer~\citep{PDETrans_ICML2025} reports a normalized RMSE, which is equivalent to the relative $L_2$ error. This work reports the mean squared error (MSE) and the relative $L_2$ error (Rel-$L_2$); the latter coincides with the metric of PDE Transformer and belongs to the same family of relative norm errors as that of ScOT and DRIFT-Net. All models are evaluated under a common protocol with shared metrics, rollout settings, mask handling, and validation-based model selection.

\section{Methods}
\label{sec:method}

\subsection{Problem Definition}

This work investigates the data-driven surrogate modeling of two-dimensional incompressible fluid flows across two problem paradigms: unsteady temporal evolutionary dynamics and stationary boundary-value mappings.

For unsteady flow systems, the dynamic progression is formulated as a first-order Markovian transition:
\begin{align}
    \hat{\mathbf{u}}_{t+\Delta t} &= \mathcal{G}_\theta(\mathbf{u}_t),
\end{align}
where $\mathbf{u}_t \in \mathbb{R}^{H \times W \times C}$ denotes the multi-channel physical state snapshot discretized on a Cartesian domain at time $t$. The parameterized surrogate $\mathcal{G}_\theta$ maps the current state to the subsequent state $\hat{\mathbf{u}}_{t+\Delta t}$ advanced by a discrete temporal increment $\Delta t$. To evaluate long-term temporal dynamics, multi-step predictions are obtained recursively via autoregressive rollout over $L$ successive horizons:
\begin{align}
    \hat{\mathbf{u}}_{t + k \Delta t} = \mathcal{G}_\theta(\hat{\mathbf{u}}_{t + (k-1)\Delta t}), \quad k = 1, 2, \dots, L,
\end{align}
initialized with the ground-truth state $\mathbf{u}_t$, such that $\hat{\mathbf{u}}_{t} = \mathbf{u}_t$.

For steady flow problems, the task is formulated as mapping spatial geometry and governing parameters to the converged steady-state field:
\begin{align}
    \hat{\mathbf{u}}_{\infty} = \mathcal{G}_\theta(\mathbf{g}, \boldsymbol{\mu}),
\end{align}
where $\mathbf{g}$ denotes the spatial domain or boundary representation, $\boldsymbol{\mu}$ represents the governing physical parameters such as the Reynolds number, and $\hat{\mathbf{u}}_{\infty} \in \mathbb{R}^{H \times W \times C}$ is the predicted steady-state field.

The optimization objective across both paradigms is to minimize the expected loss $\mathcal{L}$ between predictions and references:
\begin{equation}
    \min_\theta \; \mathbb{E} \left[ \mathcal{L}(\hat{\mathbf{u}}, \mathbf{u}) \right],
\end{equation}
where $(\hat{\mathbf{u}}, \mathbf{u})$ denotes $(\hat{\mathbf{u}}_{t+\Delta t}, \mathbf{u}_{t+\Delta t})$ for unsteady evolutions and $(\hat{\mathbf{u}}_{\infty}, \mathbf{u}_{\infty})$ for steady-state mappings.

\subsection{Model Architecture}
\label{subsec:model_architecture}

\begin{figure}[t]
  \centering
  \includegraphics[trim=22mm 2mm 2mm 0mm, clip, width=\linewidth]{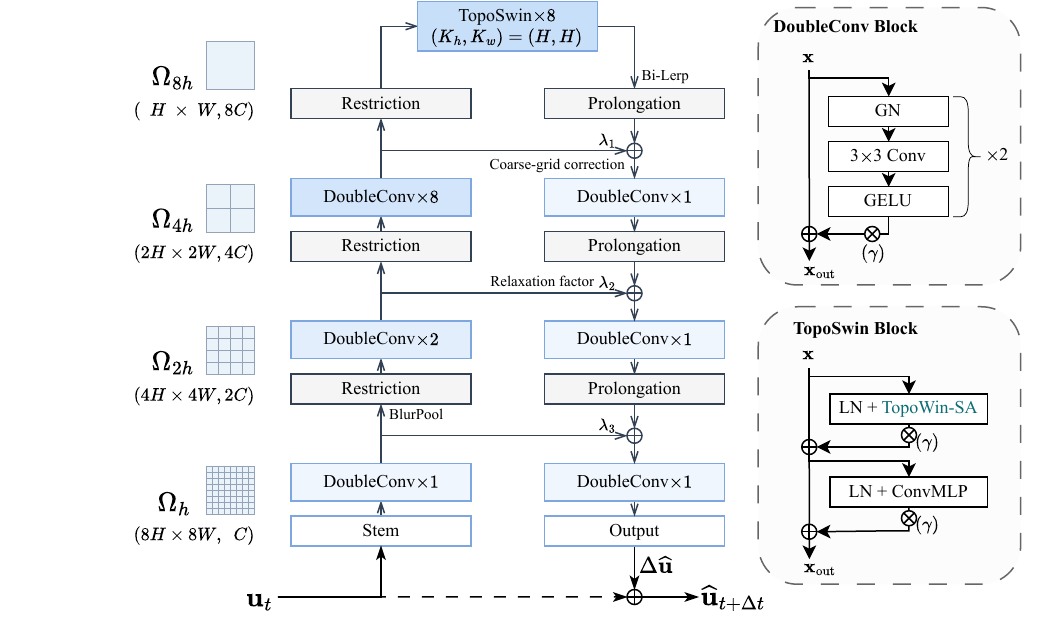}

  \caption{\textbf{Overview of the SUMNO architecture} (exemplified for $M=4$ stages). Left: multigrid-like U-shaped hierarchy across grid levels $\Omega_h \to \dots \to \Omega_{8h}$, with an asymmetric encoder--decoder, parameter-free resampling, and additive fusion with learnable relaxation factors. Right: the DoubleConv block used at fine and intermediate resolutions, and the TopoSwin block at the coarsest level, whose window attention (TopoWin-SA) follows the periodicity of each axis (highlighted). $\bigotimes(\gamma)$ denotes LayerScale. Normalization layers appear only inside sub-blocks; no normalization is applied in the inter-grid transition and fusion paths.
  See main text for details.}
  \label{fig:model_architecture}
\end{figure}

The surrogate network adopts a U-shaped multiscale topology organized across $M$ stages spanning successive spatial grid levels $\Omega_h \to \Omega_{2h} \to \dots \to \Omega_{2^{M-1}h}$. This layout resembles a single V-cycle of geometric multigrid (GMG) methods~\citep{Multigrid_Textbook2000} unrolled in space: the descent to coarser grids forms the encoder stages, and the ascent back to finer grids forms the decoder stages. An overview of the architecture (instantiated with $M=4$ stages) is illustrated in Figure~\ref{fig:model_architecture}.

Across this hierarchy, the forward pipeline is recursively structured. On the downward path for stage $m \in \{1, \dots, M-1\}$, the encoding stage first applies a DoubleConv block, analogous to pre-smoothing, followed by a restriction operator mapping the latent fields to the next coarser grid $\Omega_{2^m h}$. At the coarsest level $\Omega_{2^{M-1} h}$, a bottleneck module, analogous to a coarse-grid solver, processes domain-wide interactions. On the upward path, for stage $m$, a prolongation operator projects coarse representations back to the fine grid $\Omega_{2^{m-1}h}$, where the resulting coarse-grid corrections are combined with the lateral feature fields via fusion blocks, prior to the decoding stage, whose DoubleConv block is analogous to post-smoothing. At the finest grid level $\Omega_h$, an input stem and an output projection layer map the physical state fields to and from the latent variable space.

Optionally, for time-dependent problems, a global identity shortcut (indicated by the dashed path in Figure~\ref{fig:model_architecture}) can bypass the multiscale hierarchy to perform a discrete forward Euler update:
\begin{equation}
    \hat{\mathbf{u}}_{t+\Delta t} = \mathbf{u}_t + \Delta \hat{\mathbf{u}},
\end{equation}
where the backbone network directly predicts the temporal increment $\Delta \hat{\mathbf{u}}$. Otherwise, the output projection layer directly yields the target physical fields.

The individual operators constituting this architecture are detailed in the following sub-sections.

\subsubsection{DownSampling and UpSampling Operators}
For down-sampling, a Full Weighting (FW) type restriction operator $R: \Omega_h \to \Omega_{2h}$ is adopted, the nine-point stencil that is the scaled adjoint of bilinear interpolation. In signal processing terms, it is the low-pass anti-aliasing filter of BlurPool~\citep{BlurPool_ICML2019}, built from the three-tap binomial filter applied along each axis, and is implemented as a fixed depthwise convolution with stride 2 and the kernel
\begin{align}
    K_\text{FW}=\frac{1}{16}
\begin{bmatrix}
    1&  2&  1 \\
    2&  4&  2 \\
    1&  2&  1
\end{bmatrix}.
\end{align}
The kernel is registered as a non-trainable buffer. Circular padding is used on periodic axes, so that the stencil wraps across the domain border, and reflection padding on non-periodic axes.

For up-sampling, a bilinear interpolation prolongation operator $P: \Omega_{2h} \to \Omega_h$ is used, implemented as parameter-free cell-centered bilinear interpolation with scale factor 2. Border handling follows a pad-interpolate-crop scheme: circular padding on periodic axes, so that interpolation wraps across the border, and replicate padding on non-periodic axes.

When latent channel dimensions vary across grid levels, a pointwise ($1\times 1$) convolution is applied prior to the spatial inter-grid transfer operators; it acts on channels only, and the spatial operators $R$ and $P$ remain parameter-free. No normalization layer is applied in these transition stages.

\subsubsection{Encoder and Decoder Blocks}
Feature representations at each spatial scale are processed by Double Convolution (DoubleConv) blocks, analogous to multigrid pre- and post-smoothing operators. As illustrated in Figure~\ref{fig:model_architecture}, each DoubleConv block encapsulates two consecutive convolutional units wrapped by a single identity residual connection. A convolutional unit is defined as:
\begin{equation}
    \mathcal{T}_{\mathrm{conv}}(\cdot) = \mathrm{GELU}\Big(\mathrm{Conv}_{3\times 3}\big(\mathrm{GN}(\cdot)\big)\Big),
\end{equation}
incorporating Group Normalization (GN)~\citep{GroupNorm_ECCV2018}, a standard $3\times 3$ convolution, and a GELU~\citep{GELU_2016} activation. The residual branch is scaled by LayerScale~\citep{LayerScale_ICCV2021}, yielding the overall DoubleConv output:
\begin{equation}
    \mathbf{x}_{\text{out}} = \mathbf{x} + \boldsymbol{\gamma} \odot (\mathcal{T}_{\mathrm{conv}} \circ \mathcal{T}_{\mathrm{conv}})(\mathbf{x}),
\end{equation}
where $\boldsymbol{\gamma} \in \mathbb{R}^C$ is a learnable channel-wise scaling vector initialized to $10^{-6}$, and $\odot$ denotes element-wise multiplication along the channel dimension.

\subsubsection{Bottleneck Transformer}
At the coarsest grid level $\Omega_{2^{M-1}h}$, a stack of shifted-window Transformer blocks~\citep{Swin_ICCV2021} serves as the coarse-grid solver and couples the whole domain. Within each block, self-attention acts inside non-overlapping windows, and consecutive blocks shift the window grid by a preset offset so that neighboring windows exchange information. Periodicity is handled independently per axis by the window attention (TopoWin-SA, Section~\ref{subsec:boundary_mechanisms}); the resulting block is referred to as TopoSwin. Each TopoSwin block comprises two residual sub-layers, window self-attention and an MLP, each preceded by Layer Normalization (LN)~\citep{LayerNorm_2016}.

The window size and shift are set automatically from the feature-map size, so that the same block can be placed at any grid level and on any aspect ratio. Given preset maximum window sizes $(K_h^{\max}, K_w^{\max})$ that divide the feature-map dimensions $(H_{\text{in}}, W_{\text{in}})$, the effective window size contracts when the grid is smaller than the preset:
\begin{equation}
    K_h = \min(K_h^{\max}, H_{\text{in}}), \quad K_w = \min(K_w^{\max}, W_{\text{in}}).
\end{equation}
The shift along an axis is disabled when that axis holds a single window, since shifting would only fragment the field:
\begin{equation}
    S_i = \begin{cases}
        S_i^{\max} \bmod K_i, & \text{if } D_i > K_i, \\
        0, & \text{if } D_i = K_i,
    \end{cases} \quad \text{for } i \in \{h, w\},
    \label{eq:shift_contraction}
\end{equation}
where $D_h = H_{\text{in}}$, $D_w = W_{\text{in}}$, and $S_i^{\max}$ is the preset shift.

Queries and keys are normalized with RMS-Norm~\citep{RMSNorm_NIPS2019} before the attention logits are computed (QK-Norm~\citep{QKNorm_ICML2023}). The MLP inserts a depthwise $3\times 3$ convolution between its two linear layers, as in~\citep{CvT_ICCV2021}:
\begin{equation}
    \mathrm{ConvMLP}(\mathbf{z}) = \mathbf{W}_2 \, \mathrm{GELU}\Big(\mathrm{DWConv}_{3\times 3}\big(\mathrm{GELU}(\mathbf{W}_1 \mathbf{z} + \mathbf{b}_1)\big)\Big) + \mathbf{b}_2.
\end{equation}
Each residual branch is scaled by a learnable LayerScale vector $\boldsymbol{\gamma}_1, \boldsymbol{\gamma}_2 \in \mathbb{R}^C$, so that a TopoSwin block computes
\begin{align}
    \mathbf{z}' &= \mathbf{z} + \boldsymbol{\gamma}_1 \odot \operatorname{TopoWin\text{-}SA}\big(\mathrm{LN}(\mathbf{z})\big), \\
    \mathbf{x}_{\text{out}} &= \mathbf{z}' + \boldsymbol{\gamma}_2 \odot \mathrm{ConvMLP}\big(\mathrm{LN}(\mathbf{z}')\big).
\end{align}

\subsubsection{Fusion Operator}
On the upward path, the coarse-grid correction $\mathbf{z}_{\text{up}}$ from the prolongation operator is added to the fine-grid field $\mathbf{z}_{\text{skip}}$, as in the correction step of multigrid. The two terms are weighted by a learnable channel-wise relaxation factor to control the variance of the sum:
\begin{align}
    \boldsymbol{\lambda} &= \sigma(\boldsymbol{\alpha}), \\
    \mathbf{z}_{\text{fused}} &= \boldsymbol{\lambda} \odot \mathbf{z}_{\text{skip}} + (\mathbf{1} - \boldsymbol{\lambda}) \odot \mathbf{z}_{\text{up}},
\end{align}
where $\boldsymbol{\alpha} \in \mathbb{R}^C$ is a learnable parameter vector initialized to zero, and $\sigma(\cdot)$ denotes the Sigmoid function, so that $\boldsymbol{\lambda} \in (0, 1)^C$. The fused field $\mathbf{z}_{\text{fused}}$ is subsequently passed to the post-smoothing decoder blocks.

\subsubsection{Stem and Output Projections}
Transitions between physical state fields and latent representations are handled by an input stem and an output projection layer. The input stem maps the discrete physical fields onto the finest latent grid $\Omega_h$, and the output projection maps the decoder output on $\Omega_h$ back to the physical state space. Both modules are implemented with convolutional layers, optionally followed by activation functions; their depth is adapted to the task.

\subsection{Topology-Aware Boundary Handling}
\label{subsec:boundary_mechanisms}

Boundary treatments follow domain topology, categorized into periodic and non-periodic regimes. Along periodic axes, spatial convolutions employ circular padding to preserve continuity across cylindrical or toroidal topologies. Along non-periodic axes, spatial convolutions employ reflection padding.

\begin{figure}[ht]
    \centering
    \includegraphics[width=1.\linewidth]{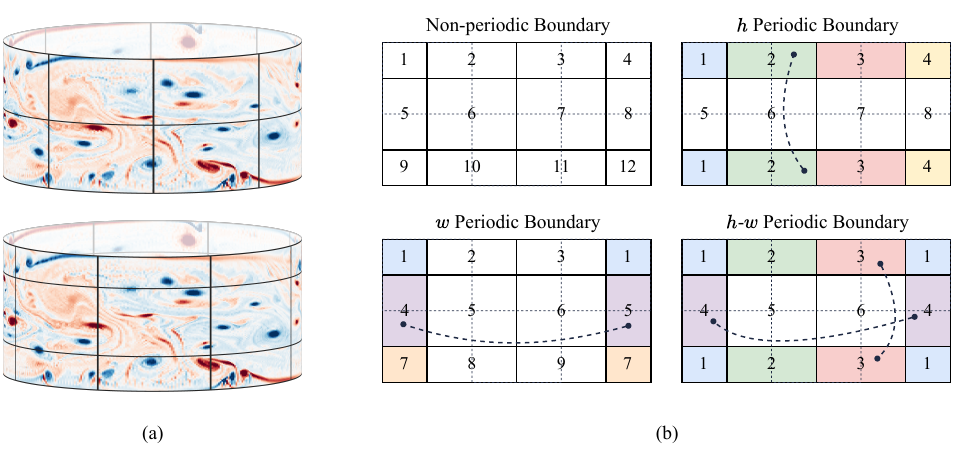}
    \caption{\textbf{Window partitioning across boundary topologies.}
    (a) Window partitioning before (top) and after (bottom) shifting for a $w$-periodic domain (cylindrical topology).
    (b) Comparison across four boundary configurations, exemplified on a 6-window grid (dashed lines). Shifting by half a window splits the 6 windows into 12 fragments when both axes are non-periodic (top left). Across periodic boundaries, fragments straddling opposing edges are merged (linked by curved dashed lines, sharing identical colors and indices), yielding 8 windows for $h$-periodicity (top right) and 9 for $w$-periodicity (bottom left), and restoring all 6 complete windows on the torus topology ($h$-$w$-periodic, bottom right).
    }
    \label{fig:tori_swin}
\end{figure}

The same distinction applies to the bottleneck window attention (TopoWin-SA). In the standard Swin Transformer~\citep{Swin_ICCV2021} processing images, a half-window shift splits the windows along both axes into fragments. For flows with a periodic direction, however, the two sides of that border are physical neighbors, so windows across it should remain intact. Figure~\ref{fig:tori_swin}(a) illustrates this on a cylindrical domain: along the periodic axis (horizontal, $w$), shifting leaves the windows intact, whereas along the non-periodic axis (vertical, $h$), fragments remain. Figure~\ref{fig:tori_swin}(b) details the resulting window behavior under all four boundary configurations. On a grid of 6 windows, a half-window shift fragments the domain into 12 sub-windows when both axes are non-periodic (top left), which is the behavior of the standard Swin Transformer. Merging the fragments across periodic borders yields 8 windows under $h$-periodicity and 9 under $w$-periodicity, and restores all 6 complete windows on a torus ($h$-$w$-periodic).

The standard Swin Transformer controls window behavior through an attention mask: tokens that a cyclic shift brings together from opposite borders are prevented from attending to each other. To obtain the window behavior described above, TopoWin-SA reformulates the mask as a single rule parameterized by the periodicity of each axis. The rule assigns each token of the shifted feature map a region tag, and a token attends only to tokens carrying the same tag. The tags are defined by an integer map $\mathbf{C} \in \{0, 1, 2, 3\}^{H_\text{in} \times W_\text{in}}$, which encodes the shifted border regions of the two axes as two bits:
\begin{equation}
    \mathbf{C}_{i, j} = \epsilon_h \, \mathbb{I}_{[H_\text{in} - S_h,\, H_\text{in})}(i) + 2\, \epsilon_w \, \mathbb{I}_{[W_\text{in} - S_w,\, W_\text{in})}(j),
\end{equation}
where $\mathbb{I}_{[\cdot)}(\cdot)$ is the interval indicator function, $(S_h, S_w)$ are the directional shift steps defined in Eq.~(\ref{eq:shift_contraction}), and $\epsilon_h, \epsilon_w \in \{0, 1\}$ equal 1 when the corresponding axis is non-periodic and 0 otherwise. For a periodic axis, $\epsilon = 0$, so tokens on both sides of its border share the same tag and attend to each other without masking.

Given $\mathbf{C}$, the attention masks are constructed per window. $\mathbf{C}$ is partitioned into windows in the same way as the tokens, and the tags of each window are compared pairwise by broadcast subtraction. Writing $\mathbf{c}_k$ for the flattened tags of the $k$-th window, the mask is
\begin{equation}
    (\mathbf{A}_k^{\text{mask}})_{p, q} =
    \begin{cases}
        0, & \text{if } (\mathbf{c}_k)_p = (\mathbf{c}_k)_q, \\
        -\infty, & \text{otherwise},
    \end{cases}
\end{equation}
where $p, q$ index the tokens within the window. When both axes are non-periodic, the mask reduces to that of the standard shifted-window attention; when both are periodic, no token pair is masked; in mixed configurations, only the non-periodic axis contributes to the mask.

\subsection{Principles of Model Configuration}
\label{subsec:model_configuration}

While Sections~\ref{subsec:model_architecture}--\ref{subsec:boundary_mechanisms} formulate the mathematical operators of individual components, this section outlines the guiding principles used to determine the architectural configurations shown in Figure~\ref{fig:model_architecture}. These choices govern two core aspects: the bottom-up determination of the grid hierarchy and stage count, and the computational-budget-driven allocation of channel widths, stage depths, and bottleneck attention windows.

\subsubsection{Hierarchical Stages via Bottleneck Resolution Limits}
In classical geometric multigrid methods, spatial coarsening is bounded by the smallest characteristic length scale of the flow field, as over-coarsening degrades coarse-grid corrections. Guided by this principle, the hierarchy is constructed bottom-up by first determining an irreducible coarse-grid dimension at the bottleneck, defined as $H \times nH$, where $n = W_{\text{in}} / H_{\text{in}}$ denotes the domain aspect ratio. Once this minimal resolution is established, the network hierarchy grows outward toward the physical input resolution through successive dyadic refinements. The required stage count $M$ is uniquely determined by the short-side input resolution $H_{\text{in}}$:
\begin{equation}
    M = \log_2\left(\frac{H_{\text{in}}}{H}\right) + 1.
\end{equation}
In the canonical configuration, the bottleneck short side is empirically set to $H = 16$. For input fields with $H_{\text{in}} = 128$, this naturally defines an $M = 4$ hierarchy across grid levels $\Omega_h \to \Omega_{2h} \to \Omega_{4h} \to \Omega_{8h}$. If higher-resolution inputs are processed, additional fine-grid stages are appended outward, preserving the same bottleneck scale.

\subsubsection{Channels, Depths and Structural Hyperparameters}
Following common practice, the number of channels doubles at each coarser level, $[C, 2C, 4C, 8C]$. For a $3\times 3$ DoubleConv block on an $H_m \times W_m = n H_m^2$ grid with $C_m$ channels, the computational cost is:
\begin{align}
    \mathrm{MACs}_{\text{DoubleConv}} = 2 \cdot 3^2  C_m^2 H_m  W_m \nonumber 
    = 18 n H_m^2 C_m^2.
\end{align}
Under dyadic scaling, the spatial area contracts fourfold ($H_{m+1} W_{m+1} = \frac{1}{4} H_m W_m$) precisely as channels double ($C_{m+1} = 2 C_m$). Consequently, the computational footprint of an individual DoubleConv block remains strictly invariant across all hierarchy levels.

This per-block cost invariance enables an asymmetric allocation of stage depths to optimize representational capacity under a constrained computational budget. Specifically, pre-smoothing depths in the encoder are progressively scaled across stages $\Omega_h \to \Omega_{4h}$ (configured as $[1, 2, 8]$ blocks with base width $C = 32$). Because the per-block cost is the same at every level, stacking blocks at coarser resolutions adds depth at the same per-block cost. Conversely, motivated by the multigrid heuristic that post-smoothing should be kept minimal to prevent excessive smoothing, the decoder depth is uniformly restricted to $1$ block per stage.

At the coarsest level $\Omega_{8h}$ ($H \times nH$ with width $8C$), a Swin Transformer block serves as the bottleneck solver. For an attention window of size $K_h \times K_w$ with an FFN expansion ratio $r = 4$, the theoretical computation per block is evaluated as:
\begin{align}
    \mathrm{MACs}_{\text{Swin}} &= (4 + 2r) H W (8C)^2 + 2 K_h K_w H W (8C) \nonumber \\
    &= 768 n H^2 C^2 + 16 n H^2 K_h K_w C.
\end{align}
Enlarging the attention window strengthens spatial representation. Under the configuration $H = 16$ and $C = 32$, setting the window dimensions to $(K_h, K_w) = (H, H) = (16, 16)$ is cheaper than increasing the number of channels, as the attention term ($16 n H^2 K_h K_w C$) remains small relative to the channel projections ($768 n H^2 C^2$). The bottleneck is configured with 8 blocks (4 regular and shifted window pairs) using these full-span $(16, 16)$ windows.

Finally, auxiliary structural hyperparameters are kept uniform across the architecture: the Transformer FFN expansion ratio is $r = 4$, convolutions use $3 \times 3$ kernels, Group Normalization is configured with $16$ groups, and the skip-connection residual scaling factors $\alpha_m$ are initialized to $0.1$.

\section{Experiments}
\label{sec:experiments}

\subsection{Benchmark Datasets}
\begin{figure*}[t]
\centering
\includegraphics[width=\textwidth]{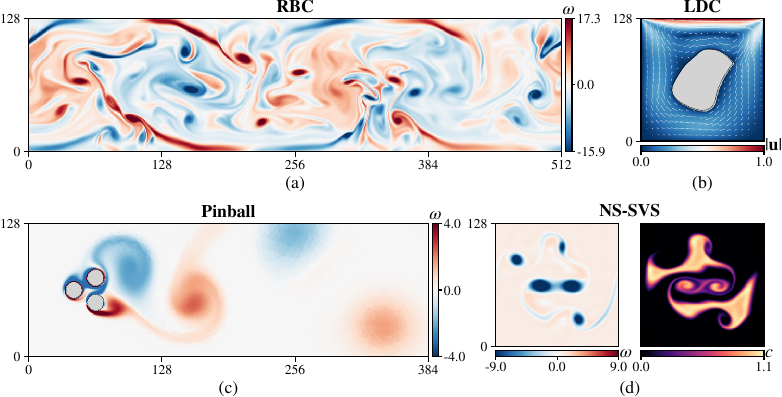}
\caption{Representative flow field snapshots selected from the four benchmark datasets: 
(a) vorticity field $\omega$ showing thermal plume structures in Rayleigh--B\'{e}nard convection (RBC, $128 \times 512$, $\mathit{Ra}=10^8$, $\mathit{Pr}=0.2$);
(b) converged velocity magnitude $|\mathbf{u}|$ overlaid with velocity vectors in the lid-driven cavity flow (LDC, $128 \times 128$, $\mathit{Re}=63$);
(c) vorticity field $\omega$ of the vortex-shedding wake behind rotating cylinders in the fluidic pinball benchmark ($128 \times 384$); 
(d) co-evolving vorticity field $\omega$ and passive scalar concentration $c$ in the Navier--Stokes sinusoidal vortex sheet problem (NS-SVS, $128 \times 128$).}
\label{fig:benchmark_visualizations}
\end{figure*}

Four representative fluid flows are investigated in this work, spanning three unsteady dynamic systems and one stationary boundary-value problem under diverse flow configurations. A comprehensive summary of governing physical variables, spatiotemporal discretizations, sample sizes, evaluation rollout horizons, and parameter partitioning protocols is provided in Table~\ref{tab:dataset_summary}, while representative flow field snapshots from each benchmark are visualized in Figure~\ref{fig:benchmark_visualizations}. The governing equations, simulation settings, and physical characteristics of each benchmark are detailed below, followed by the trajectory-level dataset splitting and parameter partitioning protocols in Section~\ref{subsec:data_processing}.

\begin{table}[ht]
\centering
\scriptsize
\caption{Benchmark configurations, target physical variables, spatiotemporal dimensions, dataset scales, evaluation rollout horizons, and parameter partitioning protocols. Windows indicates the number of training and evaluation samples (Train / Test), and Horiz. specifies the autoregressive rollout horizon during evaluation.}
\label{tab:dataset_summary}
\setlength{\tabcolsep}{3.8pt}
\renewcommand{\arraystretch}{1.15}
\begin{tabular}{@{} l @{\hspace{10pt}} l l r@{ / }l c l @{}}
\toprule
\textbf{Dataset} & \textbf{Variables} & \textbf{Grid ($H \times W,\, T$)} & \multicolumn{2}{c}{\textbf{Windows}} & \textbf{Horiz.} & \textbf{Parameter Partitioning} \\
\midrule
\textbf{NS-SVS}     & $(u, v, c)$    & $128 \times 128,\, T=21$    & 36,000 & 2,048 & 14   & Disjoint initial vorticity ($\boldsymbol{\omega}_0$) \\
\textbf{Pinball} & $(u, v, p)$    & $128 \times 384,\, T=1,001$ & 12,000 & 2,048 & 10   & Disjoint actuation controls ($\mathbf{b}$) \\
\textbf{RBC}     & $(u, v, p, b)$ & $128 \times 512,\, T=100$   & 12,000 & 2,048 & 10   & Disjoint $\mathbf{B}_0$ (Shared $Ra, Pr$) \\
\textbf{LDC}     & $(u, v, p)$    & $128 \times 128$, Steady   & 2,400  & 300   & ---  & Disjoint geometries (Shared $Re$) \\
\bottomrule
\end{tabular}
\end{table}

\begin{description}
  \item[The Navier-Stokes Sinusoidal Vortex Sheet (NS-SVS):] 
  This dataset is taken from the POSEIDON benchmark suite for fluidic foundation model \citep{ScOT_NIPS2024}. It models 2D incompressible ideal flows on a periodic torus, where initial energy is concentrated along perturbed sinusoidal curves synthesized by superimposing multiple high-frequency modes onto a fundamental wave, followed by divergence-free projection. The temporal dynamics are governed by the 2D Navier-Stokes equations coupled with a passive scalar advection equation. Each simulation trajectory is provided by the benchmark authors as uniform tensor fields across 21 temporal snapshots sampled on a discretized $128\times 128$ Cartesian grid, yielding three physical channels per state: two velocity components $(u, v)$ and the passive tracer concentration $c$.

  \item[The Actuated Fluidic Pinballs (Pinball):] 
  The dataset is derived from numerical studies on the actuated fluidic pinballs~\citep{Pinball_CSF2025}, in which three rotating cylinders are placed in a triangular configuration. The non-uniform freestream flow impinges on these bluff-body obstacles, giving rise to wake flows exhibiting the Bénard-von Kármán vortex-shedding streets. Each trajectory is parameterized by the rotation speeds and rotational directions of the three cylinders, which yield a rich variety of vortex-shedding patterns and flow dynamics. While the original high-fidelity simulations were performed on a larger computational domain with well-defined boundary conditions, a central sub-region is cropped for neural-network training, resulting in open-flow boundary conditions for the cropped dataset. The cropped subdomain is rasterized onto a uniform $128 \times 384$ Cartesian grid across all 1,001 available raw simulation snapshots, producing an open-boundary flow sequence containing velocity and pressure fields $(u,v,p)$.

  \item[Rayleigh-Bénard Convection (RBC):] 
  This dataset is adopted from the Well collection~\citep{WellRBC_NIPS2024}, where the RBC governing equations are solved via the spectral method. The simulations are conducted on a 2-D horizontally periodic domain with bottom heating. Each simulation run is governed by the non-dimensional Rayleigh number $Ra$, Prandtl number $Pr$, and initial buoyancy distribution $\mathbf{B}_0$. The published benchmark directly provides trajectories containing the velocity, pressure, and buoyancy fields $(u,v,p,b)$, formatted as uniform $128 \times 512$ Cartesian arrays spanning 200 time steps. For this study, snapshots are extracted from the latter segment ($t\in [100,199]$) of each run, where the flow has developed into a fully turbulent state.
  
  \item[Lid-Driven Cavity Flow (LDC):]
  This dataset originates from FlowBench \citep{LDC_CE2025, FlowBench_DMLR2025}, evaluating steady-state incompressible Navier--Stokes solutions within an enclosed cavity containing arbitrary internal solid obstacles. This work adopts the 2-D isothermal subset pre-rasterized on $128 \times 128$ Cartesian grids, containing the converged velocity and pressure fields $(u,v,p)$ alongside binary obstacle geometry masks across varied geometries and Reynolds numbers ($Re$). The benchmark assesses the stationary mapping from obstacle shapes to converged steady-state flow fields, conditioned on prescribed Reynolds numbers.

\end{description}

\subsection{Data Processing Protocol}
\label{subsec:data_processing}

\subsubsection{Dataset Splitting and Parameter Partitioning}

To evaluate model generalization across physical parameter configurations while eliminating temporal and spatial data leakage, all datasets are split into mutually disjoint training, validation, and test subsets at the trajectory level (or at the geometry level for the stationary LDC case). Consequently, no temporal sequences or obstacle geometries overlap across splits. An independent validation set is reserved exclusively for hyperparameter tuning and model checkpoint selection. As summarized under \textit{Parameter Partitioning} in Table~\ref{tab:dataset_summary}, the underlying parameter space is categorized into shared and disjoint components:

For NS-SVS, the initial vorticity realizations $\omega_0$ are mutually disjoint between splits. For Pinball, each subset is assigned mutually disjoint cylinder actuation controls $b_i$. For RBC, the global Rayleigh and Prandtl numbers ($Ra, Pr$) remain identically distributed across all splits, whereas trajectories originate from mutually exclusive initial buoyancy fields $\mathbf{B}_0$. For LDC, the Reynolds number $Re$ distribution is shared across splits, while the immersed obstacle geometries remains disjoint to evaluate zero-shot geometric transfer.

\subsubsection{Spatiotemporal Windowing and Autoregressive Rollout}

Rather than directly ingesting entire simulation trajectories, surrogate models are trained and evaluated on discrete temporal slices, termed \textit{spatiotemporal windows}. Formally, an individual window extracted from an underlying trajectory is characterized by its temporal origin $t_0$, sequence length $L$, and sampling stride $\Delta t_{\text{sw}}$, defining a sequence of state snapshots $\{t_0 + k \Delta t_{\text{sw}}\}_{k=0}^{L-1}$. During training, the models are optimized on short two-step windows ($L = 2$) to learn the one-step forward mapping ($t \to t + \Delta t_{\text{sw}}$). During evaluation, predictions are recursively rolled out over extended forecasting horizons ($L = \textbf{Horiz.}$ in Table~\ref{tab:dataset_summary}) to benchmark long-term performance and cumulative error propagation.

The window sampling stride is chosen on a per-dataset basis. For NS-SVS and RBC, consecutive frames are sampled using the temporal spacing as given by the dataset ($\Delta t_{\text{sw}} = \Delta t_{\text{frame}}$). For Pinball, to capture long time-horizon vortex shedding evolutions while suppressing excessive temporal redundancy across adjacent frames, windows are extracted using an interleaved stride of 10 snapshots ($\Delta t_\text{sw} = 10 \Delta t_{\text{frame}}$) initiated from staggered time offsets $t_0$. For stationary LDC, the temporal dimension degenerates ($L = 1$), and each instance corresponds to a single spatial snapshot of the converged steady-state field.

Following window extraction within each trajectory split, all valid instances are pooled and shuffled. A fixed quota is uniformly sampled from the respective pools to construct the final training and test sets (summarized under \textbf{Windows} in Table~\ref{tab:dataset_summary}).

\subsection{Baselines}
\label{subsec:baselines}

To benchmark SUMNO against current surrogate paradigms, five representative neural operator and deep learning baselines are selected across diverse architectural families. These include transformer-based operators, specifically Scalable Operator Transformer (ScOT-B)~\citep{ScOT_NIPS2024} and PDE Transformer (PDET-B)~\citep{PDETrans_ICML2025}; frequency-domain architectures, comprising DRIFT-Net-B~\citep{DRIFTNet_ICLR2026} and Fourier Neural Operator (FNO)~\citep{FNO_ICLR2021}; and the classical convolutional model U-Net~\citep{UNet_MICCAI2015}. For baseline architectures providing multiple capacity scales, variants with comparable computational budgets are chosen, denoted by the suffix ``-B''. Specifically, FNO is configured with 64 hidden channels and truncated at 32 Fourier modes along each spatial dimension.

To ensure fair and rigorous comparisons across continuous flow fields, several targeted adjustments are implemented. For the convolutional baseline, the original BatchNorm and ReLU layers are replaced with Group Normalization (using 32 groups)~\citep{GroupNorm_ECCV2018} and GELU activations~\citep{GELU_2016}, denoted as U-Net(GN). This adaptation conforms to standard SciML practices by eliminating batch-dimension statistical coupling while keeping the macro-topology identical to the original design. Additionally, because the official implementations of ScOT and DRIFT-Net only accommodate square spatial grids, minimal modifications are made to support non-square computational domains without altering their core operator designs or attention mechanics.

The parameter counts and measured computational complexities across all evaluated models are summarized in Table~\ref{tab:macs_params}. For full reproducibility, all patched baseline implementations, training scripts, and configuration files will be released in an open-source repository.

\begin{table}[ht]
\centering
\scriptsize
\caption{\textbf{Computational complexity and parameter count across evaluated models.} Multiply--accumulate operations (MACs) are measured in gigamacs (G) for a representative single forward pass, and model sizes are reported in millions (M) of parameters.}
\label{tab:macs_params}
\begin{tabular}{@{}lrr@{}}
\toprule

Models     & Measured MACs (G) & Parameters (M) \\ \midrule
ScOT-B     &                9.31&          157.63\\
PDET-B     &               19.37&          130.65\\
DRIFTNet-B &                7.74&          132.42\\
FNO        &                0.90&            8.96\\
U-Net(GN)      &               12.08&           31.04\\
Ours       &                6.31&            9.40\\

\bottomrule
\end{tabular}
\end{table}

\subsection{Evaluation Protocol and Model Selection}
For unsteady flows, both the single-step and the multi-step rollout performance are evaluated, while for the steady flow, only the final steady results are evaluated. Following prior works of ScOT and PDET, the NS-SVS and RBC are set to a rollout of 14 and 10 steps respectively. To maintain alignment, the Pinball is also evaluated on a rollout of 10 steps. For benchmarks with immersed solid bodies (i.e., the Pinball), the prediction inside the solid region is additionally reset to zero at every rollout step before being fed back as the next input, during both validation and testing. This reset is applied identically to all models, since omitting it (i.e., masking only the loss) was observed to degrade the rollout accuracy of the baselines.

Two metrics are reported for each evaluation: Mean Squared Error (MSE) and Relative $L_2$ Error (RelL2). Errors of a single sample are calculated as:
\begin{align}
l_\text{MSE}&=
\frac{\Vert \mathbf{u}-\hat{\mathbf{u}} \Vert ^2_2}{\dim (\mathbf{u})}\\
l_{\text{Rel-}L_2}&=
 \frac{\Vert \mathbf{u}-\hat{\mathbf{u}} \Vert _2}{\Vert \mathbf{u} \Vert _2},
\end{align}
where $\hat{\mathbf{u}}, \mathbf{u}\in \mathbb{R}^d$ denote the flattened prediction and ground truth fields respectively. The dimension $d$ equals to the total amount of valid fluid pixels across all channels, excluding the solid interior regions if geometric obstacle masks exist (e.g., in the Pinball and LDC benchmarks). For a whole dataset the final testing metric is the averaged error of all samples:
\begin{equation}
    \mathcal{L}_\text{test}=\frac{1}{N_\text{te}}\sum_{i=1}^{N_\text{te}} l_i,
\end{equation}
where $N_\text{te}$ is the number of test samples (windows) of a certain task and $l_i$ is the corresponding sample-wise metric of sample $i$.

During training, a multi-step rollout evaluation identical to the test protocol is periodically conducted on the validation set, whereas single-step errors are additionally logged every epoch as diagnostics only. While models are optimized using single-step transitions, the final checkpoint for each architecture is selected based on the lowest multi-step rollout error on the validation set before being evaluated on the test set.

\subsection{Training Details}
\label{subsec:training_details}

All models are trained for $300$ epochs using the AdamW optimizer \citep{AdamW_ICLR2019} with a per-step cosine annealing schedule and a global batch size of $256$. As summarized in Table~\ref{tab:hyperparameters}, training objectives follow established baseline conventions: FNO is trained with the channel-wise relative $L_2$ loss aligned with the official implementation, DRIFT-Net optimizes an $L_1$ loss combined with the Frequency-Weighted Loss (FWL)~\citep{DRIFTNet_ICLR2026}, and all other models (including SUMNO) are trained using the mean squared error (MSE). For benchmarks featuring immersed solid bodies (i.e., the Pinball and steady-state LDC datasets), training losses are computed exclusively over the fluid domain by masking out the interior obstacle region.

Key optimization hyperparameters—primarily peak learning rates, gradient clipping thresholds, and dataset-specific weight decays—are selected based on preliminary pilot runs to ensure stable optimization and proper convergence across different flow regimes. Peak learning rates and gradient clipping norms are paired with linear warmup schedules across all architectures. Weight decay is generally set to zero, as empirical sweeps indicated that non-zero $L_2$ penalties blurred fine-scale flow structures without conferring rollout stability benefits. The exceptions are FNO, which adopts its literature-standard canonical value of $1\times 10^{-4}$~\citep{FNO_ICLR2021}, and U-Net(GN), which applies $1\times 10^{-4}$ on the LDC dataset to mitigate over-parameterization. For numerical precision, most models are trained in bfloat16 (BF16), whereas FNO uses single precision (FP32) due to implementation constraints with its spectral transform operations.

For SUMNO, Group Normalization is configured with $16$ groups, and the stability constant $\epsilon$ for both normalization layers and the AdamW optimizer is set to $10^{-12}$, which was observed to improve training stability. Baseline models otherwise follow their standard implementations and default settings.

\begin{table}[t]
\centering
\scriptsize
\renewcommand{\arraystretch}{1.15}
\setlength{\tabcolsep}{3pt}
\caption{\textbf{Detailed hyperparameter configurations across models and benchmark datasets.} Dataset-specific values are ordered as (NS-SVS / Pinball / RBC / LDC). Columns list training objective (\textbf{Loss}), peak learning rate (\textbf{LR}), maximum gradient norm for clipping (\textbf{Clip}), number of linear warmup epochs (\textbf{Warmup}), weight decay (\textbf{WD}), and floating-point precision (\textbf{Dtype}).}
\label{tab:hyperparameters}
\begin{tabular*}{\columnwidth}{@{\extracolsep{\fill}} l l c c c c c @{}}
\toprule
\textbf{Model} & \textbf{Loss} & \makecell{\textbf{LR} ($\times 10^{-3}$)} & \makecell{\textbf{Clip} ($\times 10^{-2}$)} & \textbf{Warmup} & \textbf{WD} & \textbf{Dtype} \\
\midrule
ScOT     & MSE          & 0.5/1/1/1     & 0.05/0.05/0.02/0.02     & 3/6/6/50            & 0               & BF16 \\
PDET     & MSE          & 0.2/0.2/0.2/1 & 0.02/0.02/0.02/0.02     & 3/6/6/50            & 0               & BF16 \\
DRIFT-Net    & L1+FWL       & 1/2/2/1       & 0.05/0.05/0.05/0.02     & 3/6/6/50            & 0               & BF16 \\
FNO      & Rel $L_2$    & 8             & 0.30/0.30/0.30/0.10 & 3/6/6/50            & $10^{-4}$       & FP32 \\
U-Net(GN)    & MSE          & 1             & 0.05           & 3/6/6/50            & 0/0/0/$10^{-4}$ & BF16 \\
Ours     & MSE          & 1             & 0.05/0.05/0.05/0.02     & 7/14/14/50          & 0               & BF16 \\
\bottomrule
\end{tabular*}
\end{table}

\section{Results}
\label{sec:results}

\subsection{Quantitative Evaluation against Baselines}

Table~\ref{tab:main_results} presents the quantitative performance of the proposed architecture alongside established neural surrogate baselines across four fluid dynamics benchmarks, spanning both transient rollouts and steady-state fields.

\begin{table*}[t]
\centering
\scriptsize 
\setlength{\tabcolsep}{4pt} 
\renewcommand{\arraystretch}{1.17} 
\caption{\textbf{Quantitative evaluation of SUMNO and baseline architectures across physical benchmarks.} Metrics include Mean Squared Error (MSE) and Relative $L_2$ error ($\text{Rel-}L_2$) evaluated at single-step transitions (@1) and multi-step autoregressive rollout horizons (@10, @14). The best and second-best results are highlighted in \textbf{bold} and \underline{underlined}, respectively. Scale multipliers for each column are specified in parentheses.}
\label{tab:main_results}

\makebox[\linewidth][c]{
\begin{tabular}{l @{\hspace{1.5em}} ccc @{\hspace{1.5em}} ccc @{\hspace{1.5em}} c}
\toprule
\textbf{Model} 
& \textbf{SVS@1} & \textbf{Pinball@1} & \textbf{RBC@1}
& \textbf{SVS@14} & \textbf{Pinball@10} & \textbf{RBC@10}
& \textbf{LDC} \\
\midrule

\textit{MSE ($\downarrow$)} 
& \scriptsize \textit{($\times 10^{-7}$)} & \scriptsize \textit{($\times 10^{-5}$)} & \scriptsize \textit{($\times 10^{-4}$)} 
& \scriptsize \textit{($\times 10^{-5}$)} & \scriptsize \textit{($\times 10^{-3}$)} & \scriptsize \textit{($\times 10^{-3}$)} 
& \scriptsize \textit{($\times 10^{-4}$)} \\
\addlinespace[0.2em] 

ScOT-B$^\dagger$ & \underline{4.36} & 10.92 & 1.24 & 5.26 & 3.04 & 3.02 & 1.44 \\
PDET-B & 7.68 & \underline{5.24} & 1.97 & 6.08 & 1.87 & 2.29 & 1.95 \\
DRIFT-Net$^\dagger$ & 7.47 & 9.40 & 1.26 & 5.81 & 2.66 & \underline{1.93} & 15.70 \\
FNO & 4.39 & 8.89 & 4.22 & \textbf{2.35} & 2.51 & 3.31 & 57.72 \\
U-Net(GN)$^\ddagger$ & 6.93 & 6.13 & \underline{1.20} & 15.77 & \underline{1.76} & 3.06 & \textbf{0.27} \\
SUMNO & \textbf{3.35} & \textbf{2.59} & \textbf{0.60} & \underline{2.45} & \textbf{1.24} & \textbf{1.36} & \underline{0.64} \\

%
\cmidrule(lr){1-8}

\textit{Rel-$L_2$ ($\downarrow$)} 
& \scriptsize \textit{($\times 10^{-3}$)} & \scriptsize \textit{($\times 10^{-2}$)} & \scriptsize \textit{($\times 10^{-2}$)} 
& \scriptsize \textit{($\times 10^{-2}$)} & \scriptsize \textit{($\times 10^{-2}$)} & \scriptsize \textit{($\times 10^{-2}$)} 
& \scriptsize \textit{($\times 10^{-2}$)} \\
\addlinespace[0.2em] 

ScOT-B & 2.45 & 1.49 & 2.86 & 2.72 & 7.53 & 14.99 & 4.09 \\
PDET-B & 3.28 & \underline{1.05} & 3.65 & 2.99 & 6.03 & 11.59 & 5.51 \\
DRIFT-Net & 3.18 & 1.41 & 3.01 & 2.82 & 7.08 & \underline{10.57} & 7.27 \\
FNO & \underline{2.34} & 1.32 & 5.78 & \textbf{1.80} & 6.66 & 15.72 & 24.77 \\
U-Net(GN) & 3.09 & 1.07 & \underline{2.75} & 4.85 & \underline{5.56} & 15.45 & \textbf{1.92} \\
SUMNO & \textbf{2.05} & \textbf{0.71} & \textbf{1.94} & \underline{1.88} & \textbf{4.68} & \textbf{8.23} & \underline{3.54} \\

\bottomrule
\end{tabular}
}
\vspace{0.3em}
\raggedright \scriptsize $^\dagger$Original implementations only support square inputs; adaptations were introduced in the present work to accommodate non-square computational domains.
\raggedright \scriptsize $^\ddagger$GN: GroupNorm and GELU replace the BatchNorm and ReLU of the canonical U-Net.
\end{table*}

On the three unsteady benchmarks (NS-SVS, Pinball, and RBC), SUMNO achieves the lowest error in all single-step settings under both metrics. On NS-SVS, it reduces the single-step MSE by approximately $23\%$ relative to the nearest baseline (ScOT-B). On Pinball and RBC, it roughly halves the single-step MSE of the strongest baseline (PDET-B, $5.24 \to 2.59 \times 10^{-5}$; U-Net(GN), $1.20 \to 0.60 \times 10^{-4}$) and reaches single-step $\text{Rel-}L_2$ errors of $0.71 \times 10^{-2}$ and $1.94 \times 10^{-2}$, respectively.

Under extended autoregressive rollouts, SUMNO attains the lowest error on Pinball@10 ($1.24 \times 10^{-3}$ MSE, $4.68 \times 10^{-2}$ $\text{Rel-}L_2$) and RBC@10 ($1.36 \times 10^{-3}$ MSE, $8.23 \times 10^{-2}$ $\text{Rel-}L_2$). Relative to the strongest baseline in MSE, this corresponds to reductions of approximately $30\%$ on both, against U-Net(GN) on Pinball and DRIFT-Net on RBC. The three baselines with over 130M parameters (ScOT-B, PDET-B and DRIFT-Net) are all outperformed by SUMNO, with 9.4M parameters, in every setting (Table~\ref{tab:macs_params}).

SUMNO is second on the remaining two settings, NS-SVS@14 and LDC, which favor the leading baselines. On NS-SVS@14, FNO attains the lowest error ($2.35 \times 10^{-5}$ MSE, $1.80 \times 10^{-2}$ $\text{Rel-}L_2$), with SUMNO second ($2.45 \times 10^{-5}$, $1.88 \times 10^{-2}$) and the remaining baselines more than twice as high in MSE. On the steady-state lid-driven cavity (LDC) benchmark, which maps internal obstacle geometry masks to converged flow fields, U-Net(GN) attains the lowest error ($\text{Rel-}L_2 = 1.92 \times 10^{-2}$), and SUMNO is second ($3.54 \times 10^{-2}$), ahead of ScOT-B, PDET-B, DRIFT-Net and FNO.

These two settings are consistent with the strengths of the leading baselines. NS-SVS is an idealized doubly periodic flow on a torus, which matches the Fourier basis of FNO. LDC is a static mapping from obstacle geometry to a steady field rather than flow-to-flow prediction, an image-to-image task suited to convolutional U-Nets. The ranking is mirrored across the two: FNO is best on NS-SVS@14 but worst on LDC, and U-Net(GN) is best on LDC but worst on NS-SVS@14.

\subsection{Rollout Error over Steps}
\label{subsec:rollout_steps}

\begin{figure*}[ht]
\centering
\includegraphics[width=\linewidth]{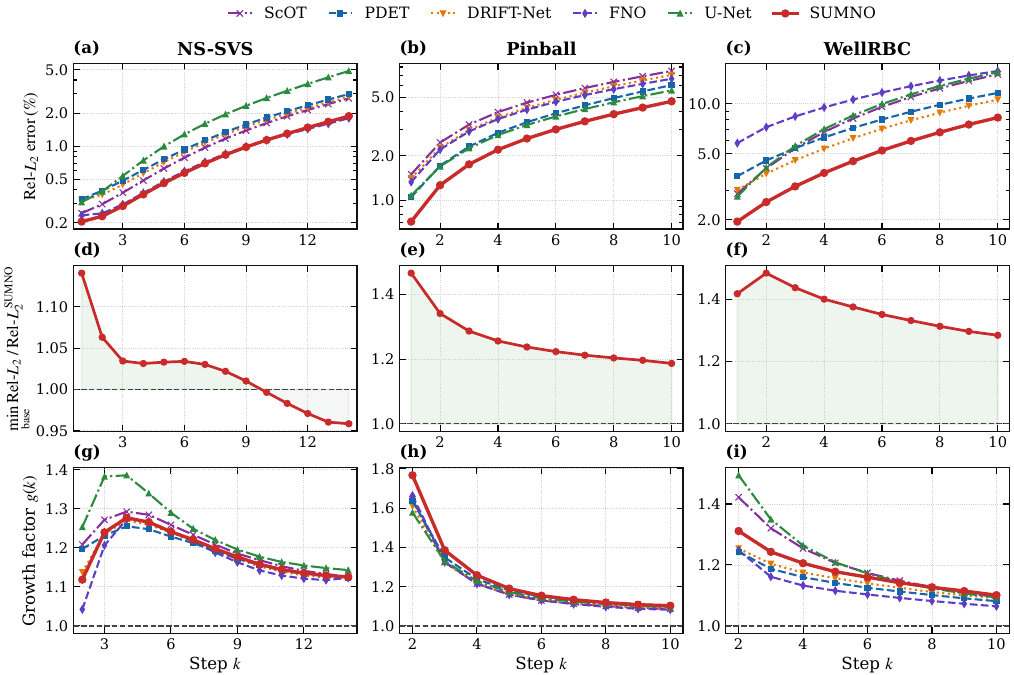}
\caption{\textbf{Rollout error analysis on the three unsteady benchmarks.}
Columns correspond to NS-SVS (14 steps), Pinball (10 steps) and RBC (10 steps); LDC is omitted because it is a steady-state mapping without rollout.
All quantities are computed from the test-set average $\text{Rel-}L_2$ at each step, and the six models share one legend.
(a--c)~$\text{Rel-}L_2$ error (in \%) at each rollout step, on a logarithmic axis.
(d--f)~Ratio of the lowest $\text{Rel-}L_2$ among the five baselines at that step to that of SUMNO; values above 1 indicate lower error for SUMNO, and the dashed line marks a ratio of 1 (equal error).
(g--i)~Step-to-step growth factor $g(k)=\text{Rel-}L_2(k)/\text{Rel-}L_2(k-1)$ for $k\ge 2$, with the dashed line marking $g=1$ (no growth).
Axis ranges are set independently for each panel.}
\label{fig:rollout_curves}
\end{figure*}

Table~\ref{tab:main_results} reports only the first and last step of each rollout. Figure~\ref{fig:rollout_curves} shows the test-set average $\text{Rel-}L_2$ at every step. Its top row (a--c) plots the error of all six models against the rollout step, its middle row (d--f) compares SUMNO with the best baseline at each step, and its bottom row (g--i) shows how fast each model's error grows from one step to the next.

The error of every model rises with the rollout step (Fig.~\ref{fig:rollout_curves}a--c). This follows from the protocol shared by ScOT, PDE Transformer and DRIFT-Net~\citep{ScOT_NIPS2024, PDETrans_ICML2025, DRIFTNet_ICLR2026}: models are trained on single-step transitions and then fed their own predictions at inference, so an error made at one step is inherited by the next. Strategies such as pushforward training or noise injection can reduce this growth~\citep{MessagePassing_ICLR2022, SimuPhy_PMLR2022, PDERefiner_NeurIPS2023}, but they also mask differences in what the networks themselves represent. The plain single-step protocol avoids this and directly exposes the representation ability and robustness of each architecture under the same error growth.

On Pinball and RBC, the curve of SUMNO lies below all baselines from the first step to the last, while the baselines cross one another and change rank along the rollout (Fig.~\ref{fig:rollout_curves}b,c). On NS-SVS, SUMNO and FNO are the two lowest curves, clearly below the other four, and U-Net is the highest (Fig.~\ref{fig:rollout_curves}a).

The middle row makes the comparison with the best baseline explicit. On Pinball and RBC the ratio stays above 1 at every step, so SUMNO leads throughout (Fig.~\ref{fig:rollout_curves}e,f). The lead is largest at the first step, where the best baseline has about 1.4 to 1.5 times the error of SUMNO, and narrows to 1.19 on Pinball and 1.28 on RBC at the tenth step. On NS-SVS the ratio crosses 1 at the tenth step: SUMNO is lower than FNO for the first nine steps, and FNO is lower afterwards, ending about 4\% below SUMNO at the fourteenth step (Fig.~\ref{fig:rollout_curves}d). The two are close over the whole rollout, and averaged over all fourteen steps they differ by less than 1\%.

The bottom row shows why the gaps are mostly set early. The growth factor of every model is largest in the first few steps and then settles at about 1.1 per step, with the curves of all models nearly overlapping late in the rollout (Fig.~\ref{fig:rollout_curves}g--i). After the first few steps, all models therefore carry their accumulated error forward at similar rates, and the gap between them changes only slowly. SUMNO's growth in the early steps is not lower than that of the baselines (on Pinball it is the highest of the six models at the second step), so its lead on Pinball and RBC originates in a lower error at the start of the rollout, which is then carried forward. On NS-SVS the late growth factors of SUMNO and FNO are practically identical, and FNO's crossing comes from its slower growth in the first steps.

\subsection{Qualitative Comparison of Predicted Flow Fields}
\label{subsec:qualitative_pred_fields}

\begin{figure*}[ht]
\centering
\includegraphics[width=\linewidth]{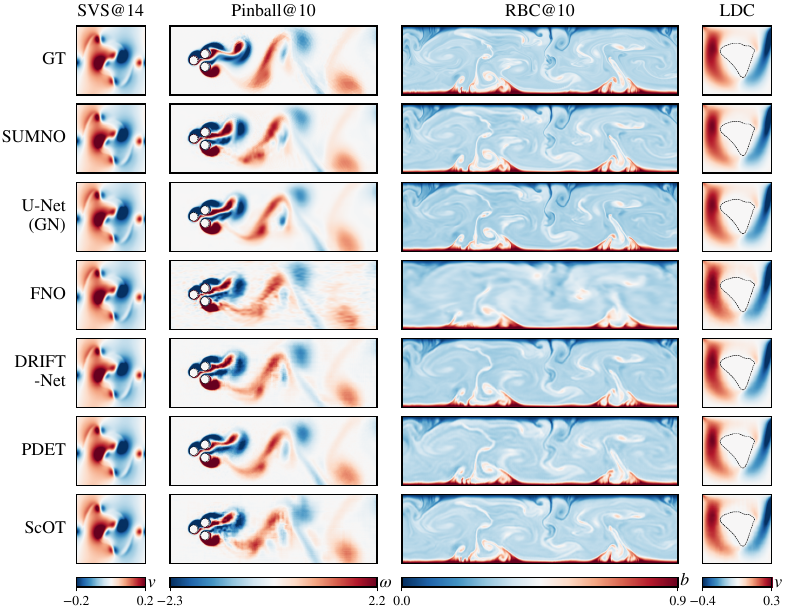}
\caption{\textbf{Qualitative comparison of predicted flow fields across the four benchmarks.} 
Columns display the velocity component $v$ for SVS@14, the vorticity field $\omega$ for Pinball@10, the buoyancy field $b$ for RBC@10, and the velocity component $v$ for steady-state LDC. 
Rows present the Ground Truth (GT) alongside predictions from SUMNO and five baseline models. 
}
\label{fig:qualitative_comparison}
\end{figure*}

Figure~\ref{fig:qualitative_comparison} compares the predicted flow fields of all models with the reference solution on a sampled test instance from each benchmark. Differences among the architectures are most noticeable on Pinball@10 and RBC@10, whereas on SVS@14 and LDC all models reproduce the main structures with little visible deviation.

On Pinball@10, evaluating the derived vorticity field $\omega = \partial_x v - \partial_y u$ exposes fine spatial differences. Under multi-step autoregressive rollouts, ScOT exhibits grid-like textures and fragmented wake patterns across the domain, and DRIFT-Net shows fainter grid-like textures in the wake. FNO displays fine-scale ripples of alternating sign that span the entire wake region. U-Net(GN) and PDET capture the global vortex layout with smooth interfaces between positive and negative vorticity regions, but over-smooth the intensity variations inside individual vortices. SUMNO retains the elongated shear layers and distinct downstream vortex cores; recurring multi-step errors nevertheless introduce localized noise and a slight bifurcation near the middle vortex core. 
All models, including the proposed one, show a slight phase lag: the leading vortex core has not advanced as far downstream as that in the ground truth. This lag can be related to training on single-step transitions while inference is autoregressive.

The buoyancy field of turbulent Rayleigh--Bénard convection (RBC@10), with heating at the bottom and cooling at the top, features a thin high-buoyancy layer (dark red) along the bottom boundary, from which slender plume stems rise into the interior and curl at their tips into mushroom-like heads. Between the stems, the interior is filled with thin, continuous filaments that wind and bend repeatedly. FNO is the blurriest: it smooths the field strongly and does not resolve the slender plume stems, which blend into broad, diffuse regions. DRIFT-Net, PDET and ScOT are visibly similar to one another, recovering the main convective fronts and plume locations while blurring the thin filaments to a comparable degree. U-Net(GN) is sharper than these three and keeps continuous contours, but some of the finest filaments are still smeared out. SUMNO preserves more of the fine filaments and their winding paths than all baselines.

On SVS@14 and the steady-state LDC benchmark, visible differences are small: all models reproduce the primary velocity structures, and in LDC all models recover the global recirculation pattern.

\subsection{Qualitative Comparison of Spatial Error Distributions}

\begin{figure*}[ht]
\centering
\includegraphics[width=\linewidth]{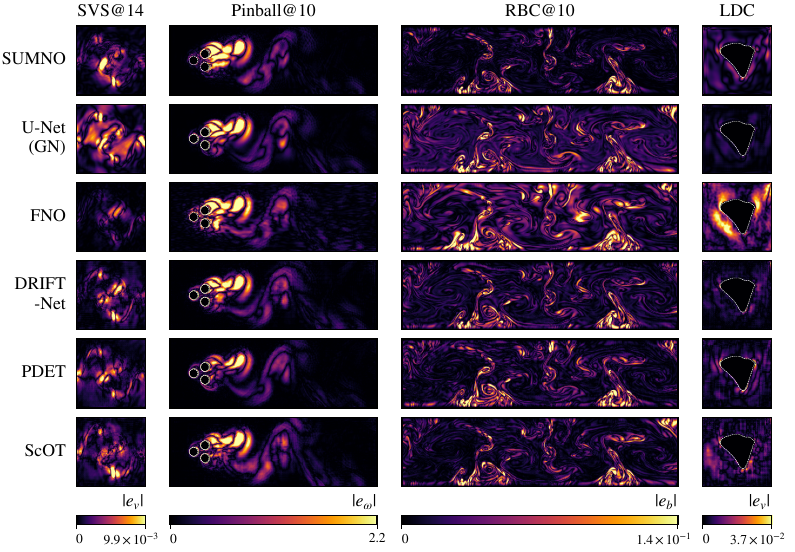}
\caption{\textbf{Point-wise absolute error distributions across the four benchmarks.}
Columns display the absolute error fields $|e_v|$ for SVS@14, $|e_\omega|$ for Pinball@10, $|e_b|$ for RBC@10, and $|e_v|$ for steady-state LDC, computed on the same sampled test instances as Figure~\ref{fig:qualitative_comparison}.
Rows compare SUMNO against five baseline architectures.
For each benchmark, the color scale is shared across all models and capped at the largest per-model 98th-percentile error.}
\label{fig:error_comparison}
\end{figure*}

Figure~\ref{fig:error_comparison} shows the point-wise absolute errors of the same test instances as Figure~\ref{fig:qualitative_comparison}. As single samples, their visual error levels may differ from the test-set averages in Table~\ref{tab:main_results}. The color scale is shared across models within each benchmark but not across benchmarks.

On SVS@14, the color scale spans only $0$ to $9.0 \times 10^{-3}$, so absolute errors are small for all models. U-Net shows elevated error across the entire field. ScOT-B, PDET-B and DRIFT-Net show similar, scattered bright patches over the domain. SUMNO shows large bright patches concentrated at the vortex cores, which suggests a misalignment of core positions. FNO has the dimmest error map overall and also locates the vortex cores accurately.

On Pinball@10, all models show large bright regions around the first shed vortex behind the cylinders, which is the phase lag discussed in Section~\ref{subsec:qualitative_pred_fields}, whereas the downstream wake is comparatively dark.
The absolute error scale ($0$ to $2.2$) matches the range of the vorticity field itself, so phase-matching errors in this region are of the same order as the signal. Together with the SVS@14 results, this suggests that the models differ in the intensity of the error, while phase matching is difficult for all of them under single-step training and autoregressive inference.

On RBC@10, bright regions follow the plumes and filaments, in contrast to the extended bright patches of the other benchmarks. FNO is brightest over the largest area, and U-Net is bright across the whole field, both along and between the filaments. ScOT-B, PDET-B and DRIFT-Net show weaker error than these two. The error of SUMNO is confined mainly to the filaments, with the regions between them close to dark, and both lateral borders also stay dark. No phase lag is evident in this sample.

On LDC, FNO has the brightest error, with large bright clumps spreading across the field and concentrated around the internal obstacle. U-Net and DRIFT-Net show dim error maps overall. ScOT-B, PDET-B and DRIFT-Net show grid-like fabric textures in their errors, whereas SUMNO shows a smooth, blurred error distribution at an overall level comparable to ScOT-B.

\section{Conclusion}
\label{sec:conclusion}

This paper presented SUMNO, a multigrid neural operator for fluid flow fields on regular grids, which spatially unrolls a multigrid V-cycle into a feed-forward U-shaped network. Its resampling, skip fusion, and normalization each follow a multigrid counterpart, hierarchically integrating convolutional processing at fine and intermediate resolutions with a shifted-window attention bottleneck at the coarsest scale. The window partitioning is generalized to handle periodicity independently per spatial axis, allowing doubly periodic, cylindrical, and non-periodic domains to be modeled within a unified rule. Evaluated across four unsteady and steady benchmarks under held-out physical configurations, SUMNO ranks first in all four settings on the complex Pinball and RBC flows, cutting 10-step rollout MSE by about 30\% on both over the strongest baselines, and leads in 5 of 7 settings overall. Operating with just 9.4M parameters, it outperforms models with over 130M parameters (ScOT-B, PDET-B, and DRIFT-Net). The remaining two settings are second place: NS-SVS at 14 steps behind FNO, and LDC behind U-Net(GN).

Limitations remain. Predicted fields still deviate visibly from the ground truth over extended rollouts; for instance, on the cylinder wake, the predicted vortices lag behind their true positions. SUMNO is designed for regular grids, and, as with data-driven surrogates in general, it does not provide formal accuracy or stability guarantees.

Future work includes extending the approach beyond regular grids to irregular or unstructured meshes and to three-dimensional flows, and enforcing exact physical invariants such as mass or energy conservation to improve the reliability of neural surrogates in practical simulations.

\section*{Acknowledgements}
This work is supported by 
the National Natural Science Foundation of China (Grant Nos. 12672294, 52676139, 12372216, W2541002, and 52276151),
the Guangdong Basic and Applied Basic Research Foundation (Grant Nos. 2026A1515011445 and 2025A1515060008), and
the Science, Technology and Innovation Commission of Shenzhen Municipality (Grant Nos. GXWD20220818113020001, GXWD20231130201948001, JCYJ20240813104853070, and JCYJ20240813105134045).
The authors appreciate the valuable discussions and constructive suggestions with Luc R. Pastur.

\section*{Declaration of competing interest}
The authors declare that there is no conflict of interest regarding the publication of this paper.

\section*{Code Availability}
The implementation code, preprocessing scripts, configuration files and trained model checkpoints required to reproduce the results in this paper will be made publicly available upon peer-reviewed publication.

\section*{Declaration of generative AI in scientific writing }
During the preparation of this work, the authors used Doubao and ChatGPT to improve the readability and language of the manuscript. After using this tool, the authors reviewed and edited the content as needed and assume full responsibility for the content of the published article.




\bibliographystyle{elsarticle-harv} 
\bibliography{custom}






\end{document}